\documentclass[10pt,twocolumn,a4paper,amsfonts,superscriptaddress, accepted=2021-12-21]{quantumarticle}
\pdfoutput=1
\usepackage[ascii]{inputenc}
\usepackage{amsmath,amssymb,amsfonts,amsthm}
\usepackage{graphicx}
\usepackage[caption=false]{subfig}
\usepackage{enumerate}
\usepackage{braket}
\usepackage[pdftex,bookmarks=false,colorlinks=true,linkcolor=blue,
citecolor=blue,filecolor=black,urlcolor=black]{hyperref}
\usepackage{tikz}
\usetikzlibrary{intersections,shapes.arrows}
\usepackage[numbers, sort&compress]{natbib}

\DeclareMathOperator{\e}{e}

\newcommand{\R}{\mathbb{R}}
\newcommand{\C}{\mathbb{C}}

\begin{document}

\title{Real time evolution with neural-network quantum states}

\author{Irene L\'opez Guti\'errez}
\email{irene.lopez@tum.de}
\affiliation{Technische Universit\"at M\"unchen, Department of Informatics and Institute for Advanced Study, Boltzmannstra{\ss}e 3, 85748 Garching, Germany}
\affiliation{Technische Universit\"at Dresden, Institute of Scientific Computing, Zellescher Weg 12-14, 01069 Dresden, Germany}

\author{Christian B.~Mendl}
\email{christian.mendl@tum.de}
\affiliation{Technische Universit\"at M\"unchen, Department of Informatics and Institute for Advanced Study, Boltzmannstra{\ss}e 3, 85748 Garching, Germany}
\affiliation{Technische Universit\"at Dresden, Institute of Scientific Computing, Zellescher Weg 12-14, 01069 Dresden, Germany}

\date{23rd December 2021}

\begin{abstract}
A promising application of neural-network quantum states is to describe the time dynamics of many-body quantum systems. To realize this idea, we employ neural-network quantum states to approximate the implicit midpoint rule method, which preserves the symplectic form of Hamiltonian dynamics. We ensure that our complex-valued neural networks are holomorphic functions, and exploit this property to efficiently compute gradients. Application to the transverse-field Ising model on a one- and two-dimensional lattice exhibits an accuracy comparable to the stochastic configuration method proposed in [\href{https://doi.org/10.1126/science.aag2302}{Carleo and Troyer, Science 355, 602--606 (2017)}], but does not require computing the (pseudo-)inverse of a matrix.
\end{abstract}

\maketitle

\section{Introduction}
The main difficulty in simulating strongly interacting many-body quantum systems on classical computers stems from the curse of dimensionality. However, a closer examination reveals that the manifold of physical quantum many-body states occupies an exponentially small volume in the Hilbert space \cite{Poulin2011}. The challenge, then, is to find an appropriate variational ansatz which has few degrees of freedom while faithfully representing physical states.

The recent successes of artificial neural network techniques have entailed a large interest in applying them to quantum many-body systems, in particular as ansatz for the wavefunction of (strongly correlated) quantum systems \cite{CarleoTroyer2017, Nomura2017, Glasser2018, Pastori2019}. Such neural-network quantum states have the principal capability to describe systems hosting chiral topological phases \cite{Glasser2018, Kaubruegger2018, Clark2018}, or to handle large entanglement \cite{GaoDuan2017, Deng2017, Levine2019}. In view of real time evolution, this could turn out to be a considerable advantage compared to established tensor network methods \cite{WhiteFeiguin2004, Vidal2004, Daley2004, Schollwock2005, Schollwock2011}, since the increase of entanglement with time demands an exponential increase of virtual bond dimensions, thus limiting the applicability of tensor network methods to relatively short time intervals \cite{Calabrese2005, AlbaCalabrese2017}.

	While the favorable capabilities of neural-network quantum states have been investigated theoretically \cite{GaoDuan2017, Levine2019}, demonstrations of their practical feasibility for quantum time evolution are still rather sparse (but see \cite{Carleo2014, CarleoTroyer2017, SchmittHeyl2018}). The canonical Dirac-Frenkel time-dependent variational principle can be regarded as projecting the time step vector onto the tangent space of the variational manifold \cite{HairerLubichWanner2006}. Time-dependent variational Monte Carlo (tdVMC) \cite{Sorella2001, Carleo2012, CarleoTroyer2017} combines the Dirac-Frenkel principle with Monte Carlo sampling and exploits the locality of typical quantum Hamiltonians. This involves the application of the (pseudo-) inverse of a covariance matrix to evolve the variational parameters in time. However, we find that in practice tdVMC can be rather sensitive to the chosen cutoff tolerance for the pseudo-inverse, or demand a prohibitively small time step for ``deep'' neural-network quantum states. Here we propose and explore an alternative approach, namely directly approximating a time step of a conventional ordinary differential equation (ODE) method by ``training'' the neural-network quantum state at the next time step using (variations of) stochastic gradient descent.

\section{Time evolution method}
Our goal is to solve the time-dependent Schr\"odinger equation
\begin{equation}\label{TDSE}
    i \frac{\partial \psi}{\partial t} = H\psi.
\end{equation}
We denote the variational ansatz by $\psi[\theta]$, where $\theta \in \C^K$ is a complex vector containing all variational parameters, which are assumed to be time-dependent. From this perspective, it is possible to find the gradients of $\psi$ with respect to $\theta$ and use the chain rule together with tangent space projections to derive an ODE for $\theta$. In stochastic reconfiguration (SR) \cite{Carleo2012, CarleoTroyer2017}, which is based on the Dirac-Frenkel variational principle, the final equation to be solved reads
\begin{equation}\label{eq:Stoch_rec}
    S \dot{ \theta } = -i F,
\end{equation}
with the covariance matrix
\begin{equation}
     S_{j,k} = \langle O_j^*; O_k\rangle
\end{equation}
and force vector
\begin{equation}
    F_j = \langle E_{\text{loc}}; O_j^* \rangle,
\end{equation}
where $\langle A; B \rangle = \langle A B \rangle - \langle A \rangle \langle B \rangle$ is the connected correlation function. The brackets denote the Monte Carlo average over samples drawn from the probability distribution $\lvert \psi[\theta] \rvert^2$, since for large system sizes it is not possible to consider the full wavefunction. For each sample $\sigma$,
\begin{align}
    O_j(\sigma) &= \partial_{\theta_j} \log(\psi[\theta](\sigma)), \\
    E_{\text{loc}}(\sigma) &= \frac{(H \psi[\theta])(\sigma)}{\psi[\theta](\sigma)}.
\end{align}
The covariance matrix $S$ is almost always effectively singular, which means it is not possible to exactly solve Eq.~\eqref{eq:Stoch_rec}. Instead, the best update for the parameters should be found by minimising $\Vert S \dot{ \theta } + i F \Vert$. One way of achieving this is by means of the Moore-Penrose pseudo-inverse. However, finding the appropriate pseudo-inverse requires choosing the right cutoff for small singular values, which can be rather challenging, specially for the real-time evolution. Krylov subspace methods, such as the conjugate gradient method or the MINRES algorithm, avoid this sensitivity problem by iteratively approximating the solution of the linear system, and have the added advantage of having low memory requirements. Their convergence to the optimal solution, however, is not guaranteed \cite{Liesen2004, Hochbruck1998}. Further note that Eq.~\eqref{eq:Stoch_rec} has dimension $K$ (number of variational parameters), which limits the feasible Ans\"atze in practice. Recently, a regularization scheme has been developed \cite{SchmittHeyl2019} which avoids issues associated with the pseudo-inverse; nevertheless, this scheme still requires to diagonalize $S$, incurring an $\mathcal{O}(K^3)$ computational cost. In general, the sensitivity of Eq.~\eqref{eq:Stoch_rec} with respect to $S$ might require a very large number of samples.

Here we propose a different approach, which fits more directly to the paradigm of neural network training: for each time step $\Delta t$, we optimize the network parameters to minimize the error
\begin{equation}\label{minimization}
    \left\Vert \psi[\theta_{n+1}] - \Phi^{\Delta t}\left(\psi[\theta_{n}]\right) \right\Vert
\end{equation}
with respect to $\theta_{n+1}$, where $\Phi^{\Delta t}$ is the discrete flow of a numerical ODE method applied to the Schr\"odinger equation. We will use the implicit midpoint method here. For an ODE $y'(t) = f(t, y(t))$ and a time-step $\Delta t$, this method is defined by
\begin{equation}
    y_{n + 1} = y_n + \Delta t f \left( t_n + \frac{\Delta t}{2}, \frac{1}{2} \left( y_n + y_{n+1} \right) \right).
\end{equation}
In the specific case of the Schr\"odinger equation, this leads to (cf.\ the Cayley transform)
\begin{equation}\label{midpoint}
    \psi[\theta_{n+1}] \approx \psi[\theta_{n}] - i \Delta t H \left(\frac{\psi[\theta_{n+1}] + \psi[\theta_{n}]}{2}\right).
\end{equation}
The implicit midpoint method has two important favorable properties: firstly, it preserves the symplectic form of Hamiltonian dynamics \cite{HairerLubichWanner2006}, and secondly, it does not contain intermediate quantities that would complicate the network optimization.

In the case of larger systems, where sampling becomes necessary, we minimize the following cost function for a single midpoint rule time step:
\begin{multline}\label{cost_sampling}
C(\theta_{n+1}) = \sum_{j=1}^N \Big\lvert \Big(\left(I + \tfrac{i \Delta t}{2} H\right)\psi[\theta_{n+1}] \\ - \left(I - \tfrac{i \Delta t}{2} H\right)\psi[\theta_n]\Big)\big(\sigma^{(j)}\big) \Big\rvert^2,
\end{multline}
with the $\sigma^{(j)}$, $j = 1, \dots, N$ a batch of input configurations, and the network parameters at the current time point, $\theta_n$, regarded as fixed. To be specific, we consider spin variables as input in the following, and denote the system size by $L$, i.e., the quantum Hilbert space dimension (number of possible spin configurations) is $2^L$. The cost function can be compactly represented in least squares form as
\begin{equation}\label{cost_function}
C(\theta) = \left\lVert A \psi[\theta] - b \right\rVert^2,
\end{equation}
with $A = \C^{N \times 2^L}$ the (sparse) submatrix of $I + \tfrac{i \Delta t}{2} H$ containing the rows corresponding to the spin configurations $\sigma^{(j)}$, and the vector $b \in \C^N$ with entries
\begin{equation}
b_j = \left(\left(I - \tfrac{i \Delta t}{2} H\right)\psi[\theta_n]\right)\big(\sigma^{(j)}\big).
\end{equation}

Assuming that $\psi[\theta]$ is a holomorphic function of the parameters $\theta$ and following the derivation in appendix \ref{sec:Wirtinger} leads to
\begin{equation}
\frac{\partial C(\theta)}{\partial \theta_{\ell}} = \left\langle A \psi[\theta] - b \Big\vert A \frac{\partial \psi[\theta]}{\partial \theta_{\ell}} \right\rangle.
\end{equation}
Note that since $C(\theta)$ is not holomorphic, the partial derivative on the left side of this expression is a Wirtinger derivative.

Appendix~\ref{sec:Geometrical} provides a geometrical perspective on the SR and implicit midpoint methods.

\section{Application to the Ising chain}
\begin{figure}[!ht]
\subfloat[Restricted Boltzmann machine architecture]{
    \begin{tikzpicture}
    \draw [thin] (0,2.5) -- (0.5,0);
    \draw [thin] (2,2.5) -- (0.5,0);
    \draw [thin] (4,2.5) -- (0.5,0);
    \draw [thin] (6.5,2.5) -- (0.5,0);
    \draw [thin] (0,2.5) -- (2.5,0);
    \draw [thin] (2,2.5) -- (2.5,0);
    \draw [thin] (4,2.5) -- (2.5,0);
    \draw [thin] (6.5,2.5) -- (2.5,0);
    \draw [thin] (0,2.5) -- (6,0);
    \draw [thin] (2,2.5) -- (6,0);
    \draw [thin] (4,2.5) -- (6,0);
    \draw [thin] (6.5,2.5) -- (6,0);
    \draw [color=black, fill=orange!50] (0.5,0) circle [radius=0.5];
    \draw [color=black, fill=orange!50] (2.5,0) circle [radius=0.5];
    \draw [dotted, ultra thick] (4,0) -- (4.5,0);
    \draw [color=black, fill=orange!50] (6,0) circle   [radius=0.5];
    \draw [color=black, fill=blue!30] (0,2.5) circle   [radius=0.5];
    \draw [color=black, fill=blue!30] (2,2.5) circle   [radius=0.5];
    \draw [color=black, fill=blue!30] (4,2.5) circle   [radius=0.5];
    \draw [dotted, ultra thick] (5,2.5) -- (5.5,2.5);
    \draw [color=black, fill=blue!30] (6.5,2.5) circle [radius=0.5];
    \node at (0.5,0) {$\sigma_1$};
    \node at (2.5,0) {$\sigma_2$};
    \node at (6,0) {$\sigma_L$};
    \node at (0,2.5) {$h_1$};
    \node at (2,2.5) {$h_2$};
    \node at (4,2.5) {$h_3$};
    \node at (6.5,2.5) {$h_M$};
    \end{tikzpicture}
    \label{RBM}}\\
\subfloat[Time evolution error]{
    \includegraphics[width=1\columnwidth]{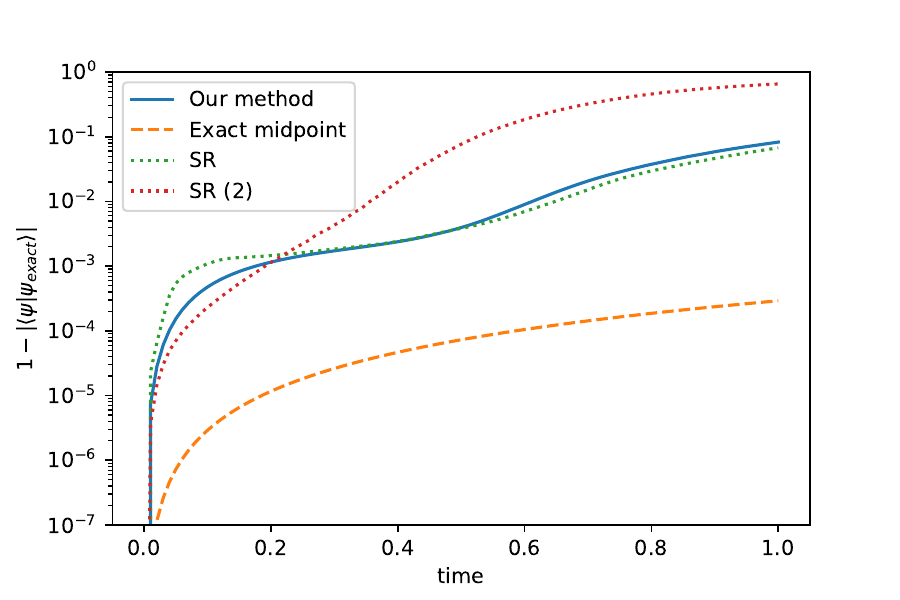}
    \label{fig:comparison}}
\caption{(a) Schematic diagram of a restricted Boltzmann machine showing the visible (orange) and hidden (blue) layers and the connections between the neurons. (b) Error in the time evolution performed using a RBM with 80 hidden units (1700 complex parameters in total), time step 0.01 and 50,000 uniformly drawn samples per time step. For comparison, SR and SR (2) show the stochastic reconfiguration method proposed in \cite{CarleoTroyer2017} with two different pseudo-inverse thresholds.}
\end{figure}
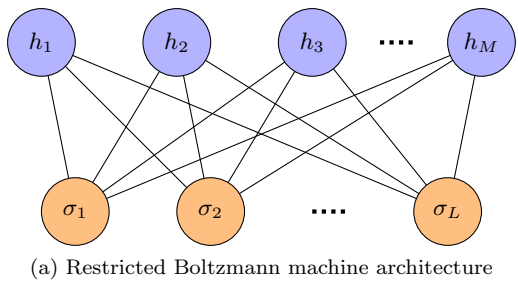
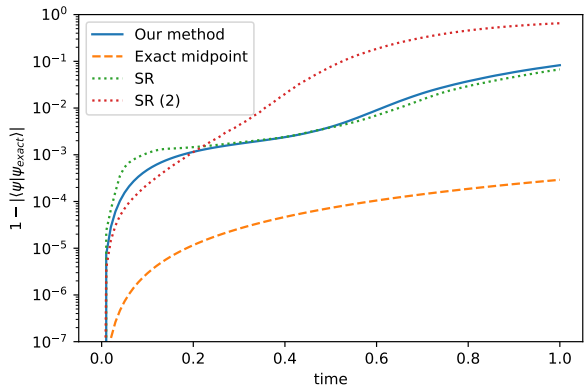

The neural-network architecture proposed as a wavefunction ansatz in \cite{CarleoTroyer2017} is the restricted Boltzmann machine (RBM). It has been shown to represent the ground state of various Hamiltonians with high accuracy \cite{Borin2020}. As visualized in Fig.~\ref{RBM}, a RBM consists of two layers of neurons, referred to as the `visible' and `hidden' layers, which are connected with one another but have no intra-layer connections. The input of the visible layer is a specific spin configuration $\sigma$. From this architecture one obtains a variational ansatz for the wavefunction of the system
\begin{equation}\label{eq:RBM_wavefunction}
    \psi(\sigma) = \sum_{\{h_i\}} \e^{\sum_j a_j \sigma_j + \sum_i b_i h_i + \sum_{i,j} w_{ij} h_i \sigma_j},
\end{equation}
where $a_j, b_i$ and $w_{ij}$ are the network parameters, and $h_i$ are the auxiliary spin variables which only take the values $\pm 1$. Due to the sumation over $h_i$ in Eq.~\eqref{eq:RBM_wavefunction}, it is possible to trace out the hidden spin variables:
\begin{equation}
    \psi(\sigma) = \e^{\sum_j a_j \sigma_j} \prod_i 2 \text{cosh}\Big(b_i + \sum_j w_{ij} \sigma_j\Big).
\end{equation}

In this section, we apply the RMB ansatz to the one-dimensional transverse-field Ising model, which consists of a chain of spins that interact with their nearest neighbors and are subject to an external magnetic field, $h$. Its Hamiltonian for general lattice dimension is given by 
\begin{equation}
\label{eq:hamiltonian}
    H_{\text{TFI}} = - J \sum_{\langle i , j \rangle} \sigma_i^z \sigma_j^z - h \sum_i \sigma_i^x.
\end{equation}
Such a system undergoes a phase transition from the ferromagnetic to the paramagnetic regime at $h = J$ in one dimension \cite{Suzuki2013} and at $h/J = 3.04438(2)$ in two dimensions \cite{BloeteDeng2002}. In the following, we set the coupling constant $J$ to $1$.

We consider a chain of length $L = 20$, which is small enough to compute the exact time evolution as a reference. For the training we use the Adam optimizer and choose the recommended hyperparameters from Ref.~\cite{KingmaBa2015}.

Fig.~\ref{fig:comparison} shows a comparison between the error obtained with stochastic reconfiguration and the error of our method. The time evolution labeled SR was computed with a pseudo-inverse threshold of $10^{-10}$, while for SR(2) a threshold of $10^{-9}$ was used. The importance and sensitivity of this threshold is reflected in the difference in the final error. We also plot the contribution to the error by the midpoint method. Even without a finely tuned optimization, our method yields comparable results to stochastic reconfiguration with an optimal pseudo-inverse cut-off. The additional error compared to the exact midpoint integration hints towards a lack of expressibility resulting from the RBM ansatz.

Since we are interested in physically realistic states, the initial state was found by performing a Hamiltonian quench with respect to the field strength $h$. We first optimize the network parameters to represent the ground state for $h = 1.5$ and then change $h$ to $0.75$ for the real time evolution. As a measure of the accuracy of the ground state, the deviation from the exact energy is $|(E[\theta] - E_0)/E_0| \approx 0.051$, where $E_{0}$ is the exact energy found by diagonalising the Hamiltonian and $E[\theta]$ is the energy of the quantum neural network state.

The training was performed on a single CPU and 64~GB memory, using multi-threading on 28 cores. We have implemented the neural network architectures and optimization part of the algorithm in a custom C code, with a Python interface to schedule and evaluate the runs. In this setup we find that a single optimization step of the RBM network with 20 lattice sites and 50,000 samples takes approximately 3.6s.

\section{Application to the Ising model on a square lattice}

\begin{figure}[!b]
\begin{tikzpicture}
\node at (0.75,4) {$\sigma$};
\filldraw[color=black, fill=blue!30] (0,0) -- (1.5,1.5) -- (1.5,4.5) -- (0,3) -- cycle;

\filldraw[color=black, fill=orange!50] (0.2, 2.8) -- (0.2,2) -- (0.7,2.5) -- (0.7,3.3) -- cycle;

\draw [thin] (0.2,2.8) -- (2.1,2.4);
\draw [thin] (0.2,2) -- (2.1,2);
\draw [thin] (0.7,2.5) -- (2.3,2.2);
\draw [thin] (0.7,3.3) -- (2.3,2.6);

\filldraw[color=black, fill=white, thin] (2.1,2.4) -- (2.1,2) -- (2.3,2.2) -- (2.3,2.6) -- cycle;

\filldraw [color=black, fill=blue!10, thin] (2,0.5) -- (3,1.5) -- (3,3.5) -- (2,2.5) -- cycle;
\filldraw [color=black, fill=blue!10,thin] (2.2,0.5) -- (3.2,1.5) -- (3.2,3.5) -- (2.2,2.5) -- cycle;
\filldraw [color=black, fill=blue!10,thin] (2.4,0.5) -- (3.4,1.5) -- (3.4,3.5) -- (2.4,2.5) -- cycle;
\filldraw [color=black, fill=blue!10,thin] (2.6,0.5) -- (3.6,1.5) -- (3.6,3.5) -- (2.6,2.5) -- cycle;
\filldraw [color=black, fill=blue!10,thin] (2.8,0.5) -- (3.8,1.5) -- (3.8,3.5) -- (2.8,2.5) -- cycle;

\draw [thin] (3, 2.3) -- (5,4);
\draw [thin] (3, 2.3) -- (5,3);
\draw [thin] (3, 2.3) -- (5,0.5);

\draw [thin] (5, 4) -- (6.5,2.2);
\draw [thin] (5, 3) -- (6.5,2.2);
\draw [thin] (5, 0.5) -- (6.5,2.2);

\filldraw [fill=orange!50] (5, 4) circle [radius=0.3];
\filldraw [fill=orange!50] (5, 3) circle [radius=0.3];
\filldraw [fill=orange!50] (5,0.5) circle [radius=0.3];
\draw [dotted, ultra thick] (5, 2.1) -- (5,1.4);

\filldraw [fill=orange!50] (6.5, 2.2) circle [radius=0.3];
\node at (7.5,2.2) {$\text{CNN}(\sigma)$};
\end{tikzpicture}
\caption{Schematic diagram of a convolutional neural network.}
\label{fig:CNN}
\end{figure}
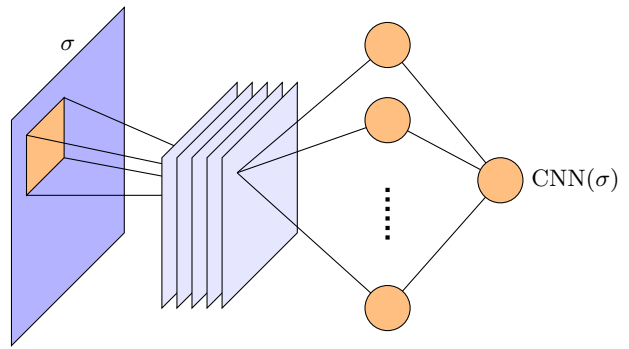

\begin{figure*}
\centering
\includegraphics[width=0.9\textwidth]{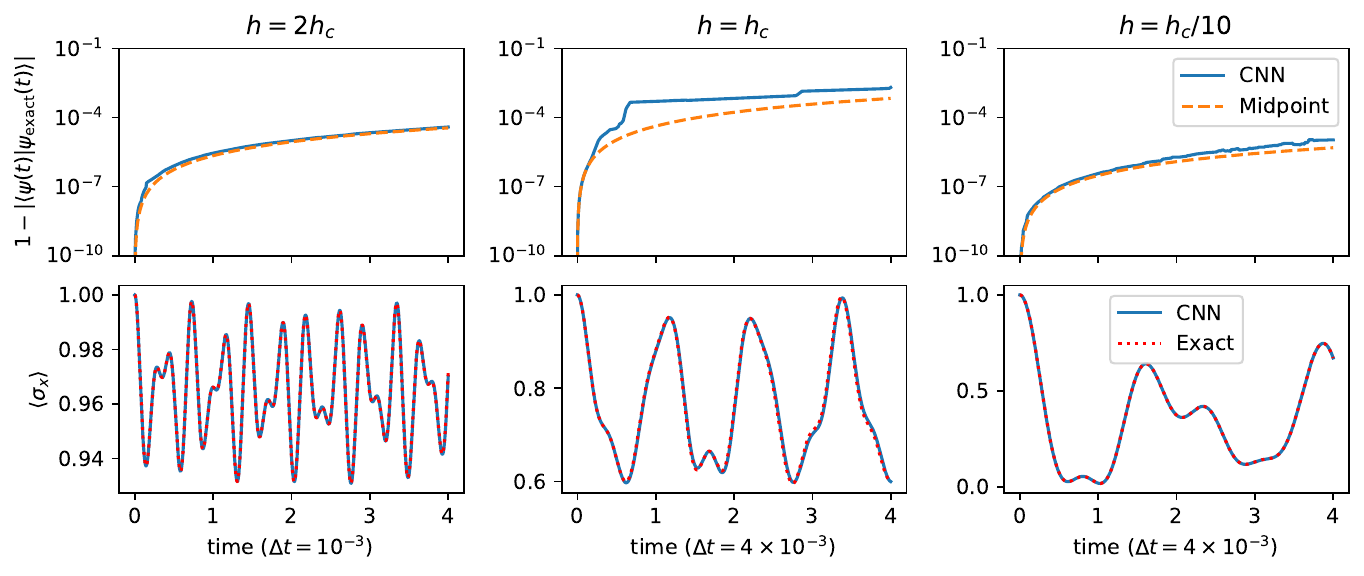}
\caption{Overlap error (top row) and transverse magnetization (bottom row) of the real time evolution governed by the Ising Hamiltonian \eqref{eq:hamiltonian} on a $3 \times 3$ lattice after a quench of $h$. Each column corresponds to a different value of $h$ after the quench, starting from ``infinite'' $h$. We used 500 uniformly drawn samples for each individual optimization step of our method (CNN).}
\label{fig:2d_tevol}
\end{figure*}

To demonstrate the flexibility of our method we change the neural network architecture and consider the time evolution governed by the two-dimensional Ising model on a $L \times L$ lattice with periodic boundary conditions, setting $L=3$ in the following. As an ansatz, we directly use the output of a convolutional neural network,
\begin{equation}
\psi(\sigma) = \text{CNN}(\sigma)
\end{equation}
as shown in Fig.~\ref{fig:CNN}. Specifically, the network architecture consists of a convolutional layer with five complex-valued $2 \times 2$ filters and periodic boundary conditions, an intermediate dense layer with 10 neurons, and a single output. This correspond to 496 complex parameters in total. For the activation functions, we found the polynomial functions described in Ref.~\cite{SchmittHeyl2019} to achieve errors several orders of magnitude smaller compared to other more commonly used activation functions, such as ReLUs or sigmoid functions. Specifically, we use
\begin{equation}
\sigma(z) = \frac{1}{2}z^2 - \frac{1}{12}z^4 + \frac{1}{45}z^6
\end{equation}
for the convolutional layer, and
\begin{equation}
\sigma(z) = z - \frac{1}{3}z^3 + \frac{2}{15}z^5
\end{equation}
for subsequent layers.

We perform three different Hamiltonian quenches, from ``infinite'' $h$ to $h = 2h_c$, $h_c$ and $h_c / 10$, where $h_c$ is the critical point at which the system undergoes a phase transition. Here ``infinite'' $h$ is equivalent to retaining only the second term in the Hamiltonian \eqref{eq:hamiltonian}, such that the corresponding ground state is the para\-magnetic state $\psi_0 = \prod_{j=1}^L \ket{+}_j$ with $\ket{+} = \frac{1}{\sqrt{2}}(\ket{0} + \ket{1})$. We are able to precisely represent $\psi_0$ using our network ansatz, such that the relative energy error is on the order of $10^{-11}$.

The top row in Fig.~\ref{fig:2d_tevol} shows the overlap error with respect to the exact wavefunction as a function of time, as well as the error resulting from a plain midpoint integration. Each time step was optimized using the same learning rate and number of iterations, but a more careful optimization could further lower the error. The bottom row of Fig.~\ref{fig:2d_tevol} shows the evolution of the transverse magnetization, based on our CNN ansatz and the numerically exact curve as reference. In agreement with Ref.~\cite{Czischek2018}, we find quenches to the critical point to be the hardest to optimize.

\section{Conclusion}
We have demonstrated that established methods for neural network optimization can be employed to describe the real time evolution of quantum wavefunctions. One additional advantage of this method, different from SR, is that it allows for the neural network architecture to be changed on the fly, which could be useful for capturing the growing complexity of the system as time progresses.

Taking full advantage of advanced machine learning techniques could further improve the results presented here, e.g., using deeper network architectures with batch normalization and residual blocks \cite{Shrestha2019}. In this work, the network parameters were optimised without any restrictions, but imposing symmetries or a certain structure, especially to the CNN filters, could accelerate and improve the optimization.

\textbf{Acknowledgments}
We thank Jan Budich and Lorenzo Pastori for useful discussions, and the Munich Center for Quantum Science and Technology for support.

\emph{Note:} During the preparation of this manuscript we became aware of related work by Markus Schmitt and Markus Heyl \cite{SchmittHeyl2019}, which appeared simultaneously on arXiv.

\bibliographystyle{plainnat}
\bibliography{references}

\begin{thebibliography}{34}
\providecommand{\natexlab}[1]{#1}
\providecommand{\url}[1]{\texttt{#1}}
\expandafter\ifx\csname urlstyle\endcsname\relax
  \providecommand{\doi}[1]{doi: #1}\else
  \providecommand{\doi}{doi: \begingroup \urlstyle{rm}\Url}\fi

\bibitem[Alba and Calabrese(2017)]{AlbaCalabrese2017}
V.~Alba and P.~Calabrese.
\newblock {Entanglement and thermodynamics after a quantum quench in integrable
  systems}.
\newblock \emph{PNAS}, 114:\penalty0 7947--7951, 2017.
\newblock \doi{10.1073/pnas.1703516114}.

\bibitem[Bl\"ote and Deng(2002)]{BloeteDeng2002}
H.~W.~J. Bl\"ote and Y.~Deng.
\newblock {Cluster Monte Carlo simulation of the transverse Ising model}.
\newblock \emph{Phys. Rev. E}, 66:\penalty0 066110, 2002.
\newblock \doi{10.1103/PhysRevE.66.066110}.

\bibitem[Borin and Abanin(2020)]{Borin2020}
A.~Borin and D.~A. Abanin.
\newblock {Approximating power of machine-learning ansatz for quantum many-body
  states}.
\newblock \emph{Phys. Rev. B}, 101, 2020.
\newblock \doi{10.1103/PhysRevB.101.195141}.

\bibitem[Calabrese and Cardy(2005)]{Calabrese2005}
P.~Calabrese and J.~Cardy.
\newblock {Evolution of entanglement entropy in one-dimensional systems}.
\newblock \emph{J. Stat. Mech.: Theory Exp.}, 2005:\penalty0 P04010, 2005.
\newblock \doi{10.1088/1742-5468/2005/04/p04010}.

\bibitem[Carleo and Troyer(2017)]{CarleoTroyer2017}
G.~Carleo and M.~Troyer.
\newblock {Solving the quantum many-body problem with artificial neural
  networks}.
\newblock \emph{Science}, 355:\penalty0 602--606, 2017.
\newblock \doi{10.1126/science.aag2302}.

\bibitem[Carleo et~al.(2012)Carleo, Becca, Schir\'o, and Fabrizio]{Carleo2012}
G.~Carleo, F.~Becca, M.~Schir\'o, and M.~Fabrizio.
\newblock {Localization and glassy dynamics of many-body quantum systems}.
\newblock \emph{Sci. Rep.}, 2:\penalty0 243, 2012.
\newblock \doi{10.1038/srep00243}.

\bibitem[Carleo et~al.(2014)Carleo, Becca, Sanchez-Palencia, Sorella, and
  Fabrizio]{Carleo2014}
G.~Carleo, F.~Becca, L.~Sanchez-Palencia, S.~Sorella, and M.~Fabrizio.
\newblock {Light-cone effect and supersonic correlations in one- and
  two-dimensional bosonic superfluids}.
\newblock \emph{Phys. Rev. A}, 89:\penalty0 031602, 2014.
\newblock \doi{10.1103/PhysRevA.89.031602}.

\bibitem[Clark(2018)]{Clark2018}
S.~R. Clark.
\newblock {Unifying neural-network quantum states and correlator product states
  via tensor networks}.
\newblock \emph{J. Phys. A Math. Theor.}, 51:\penalty0 135301, 2018.
\newblock \doi{10.1088/1751-8121/aaaaf2}.

\bibitem[Czischek et~al.(2018)Czischek, G\"arttner, and Gasenzer]{Czischek2018}
S.~Czischek, M.~G\"arttner, and T.~Gasenzer.
\newblock {Quenches near Ising quantum criticality as a challenge for
  artificial neural networks}.
\newblock \emph{Phys. Rev. B}, 98:\penalty0 024311, 2018.
\newblock \doi{10.1103/PhysRevB.98.024311}.

\bibitem[Daley et~al.(2004)Daley, Kollath, Schollw\"ock, and Vidal]{Daley2004}
A.~J. Daley, C.~Kollath, U.~Schollw\"ock, and G.~Vidal.
\newblock {Time-dependent density-matrix renormalization-group using adaptive
  effective Hilbert spaces}.
\newblock \emph{J. Stat. Mech. Theory Exp.}, 2004:\penalty0 P04005, 2004.
\newblock \doi{10.1088/1742-5468/2004/04/p04005}.

\bibitem[Deng et~al.(2017)Deng, Li, and Das~Sarma]{Deng2017}
D.~Deng, X.~Li, and S.~Das~Sarma.
\newblock {Quantum entanglement in neural network states}.
\newblock \emph{Phys. Rev. X}, 7:\penalty0 021021, 2017.
\newblock \doi{10.1103/PhysRevX.7.021021}.

\bibitem[Gao and Duan(2017)]{GaoDuan2017}
X.~Gao and L.-M. Duan.
\newblock {Efficient representation of quantum many-body states with deep
  neural networks}.
\newblock \emph{Nat. Commun.}, 8:\penalty0 662, 2017.
\newblock \doi{10.1038/s41467-017-00705-2}.

\bibitem[Glasser et~al.(2018)Glasser, Pancotti, August, Rodriguez, and
  Cirac]{Glasser2018}
I.~Glasser, N.~Pancotti, M.~August, I.~D. Rodriguez, and J.~I. Cirac.
\newblock {Neural-network quantum states, string-bond states, and chiral
  topological states}.
\newblock \emph{Phys. Rev. X}, 8:\penalty0 011006, 2018.
\newblock \doi{10.1103/PhysRevX.8.011006}.

\bibitem[Hairer et~al.(2006)Hairer, Lubich, and Wanner]{HairerLubichWanner2006}
E.~Hairer, C.~Lubich, and G.~Wanner.
\newblock \emph{{Geometric Numerical Integration. Structure-Preserving
  Algorithms for Ordinary Differential Equations}}.
\newblock Springer-Verlag Berlin Heidelberg, 2006.
\newblock \doi{10.1007/3-540-30666-8}.

\bibitem[Hirose(2012)]{Hirose2012}
A.~Hirose.
\newblock \emph{{Complex-Valued Neural Networks}}.
\newblock Springer-Verlag Berlin Heidelberg, 2012.
\newblock \doi{10.1007/978-3-642-27632-3}.

\bibitem[Hochbruck and Lubich(1998)]{Hochbruck1998}
M.~Hochbruck and C.~Lubich.
\newblock {Error analysis of Krylov methods in a nutshell}.
\newblock \emph{SIAM J. Sci. Comput.}, 19\penalty0 (2):\penalty0 695--701,
  1998.
\newblock \doi{10.1137/S1064827595290450}.

\bibitem[Kaubruegger et~al.(2018)Kaubruegger, Pastori, and
  Budich]{Kaubruegger2018}
R.~Kaubruegger, L.~Pastori, and J.~C. Budich.
\newblock {Chiral topological phases from artificial neural networks}.
\newblock \emph{Phys. Rev. B}, 97:\penalty0 195136, 2018.
\newblock \doi{10.1103/PhysRevB.97.195136}.

\bibitem[Kingma and Ba(2015)]{KingmaBa2015}
D.~P. Kingma and J.~Ba.
\newblock {Adam: A method for stochastic optimization}.
\newblock In \emph{3rd International Conference for Learning Representations,
  San Diego}, 2015.

\bibitem[Levine et~al.(2019)Levine, Sharir, Cohen, and Shashua]{Levine2019}
Y.~Levine, O.~Sharir, N.~Cohen, and A.~Shashua.
\newblock {Quantum entanglement in deep learning architectures}.
\newblock \emph{Phys. Rev. Lett.}, 122:\penalty0 065301, 2019.
\newblock \doi{10.1103/PhysRevLett.122.065301}.

\bibitem[Liesen and Tich\'y(2004)]{Liesen2004}
J.~Liesen and P.~Tich\'y.
\newblock {Convergence analysis of Krylov subspace methods}.
\newblock \emph{GAMM-Mitteilungen}, 27:\penalty0 153--173, 2004.
\newblock \doi{10.1002/gamm.201490008}.

\bibitem[Nomura et~al.(2017)Nomura, Darmawan, Yamaji, and Imada]{Nomura2017}
Y.~Nomura, A.~S. Darmawan, Y.~Yamaji, and M.~Imada.
\newblock {Restricted Boltzmann machine learning for solving strongly
  correlated quantum systems}.
\newblock \emph{Phys. Rev. B}, 96:\penalty0 205152, 2017.
\newblock \doi{10.1103/PhysRevB.96.205152}.

\bibitem[Pastori et~al.(2019)Pastori, Kaubruegger, and Budich]{Pastori2019}
L.~Pastori, R.~Kaubruegger, and J.~C. Budich.
\newblock {Generalized transfer matrix states from artificial neural networks}.
\newblock \emph{Phys. Rev. B}, 99:\penalty0 165123, 2019.
\newblock \doi{10.1103/PhysRevB.99.165123}.

\bibitem[Poulin et~al.(2011)Poulin, Qarry, Somma, and Verstraete]{Poulin2011}
D.~Poulin, A.~Qarry, R.~Somma, and F.~Verstraete.
\newblock {Quantum simulation of time-dependent Hamiltonians and the convenient
  illusion of Hilbert space}.
\newblock \emph{Phys. Rev. Lett.}, 106:\penalty0 170501, 2011.
\newblock \doi{10.1103/PhysRevLett.106.170501}.

\bibitem[Schmitt and Heyl(2018)]{SchmittHeyl2018}
M.~Schmitt and M.~Heyl.
\newblock {Quantum dynamics in transverse-field Ising models from classical
  networks}.
\newblock \emph{SciPost Phys.}, 4:\penalty0 013, 2018.
\newblock \doi{10.21468/SciPostPhys.4.2.013}.

\bibitem[Schmitt and Heyl(2020)]{SchmittHeyl2019}
M.~Schmitt and M.~Heyl.
\newblock {Quantum many-body dynamics in two dimensions with artificial neural
  networks}.
\newblock \emph{Phys. Rev. Lett.}, 125:\penalty0 100503, 2020.
\newblock \doi{10.1103/PhysRevLett.125.100503}.

\bibitem[Schollw\"ock(2005)]{Schollwock2005}
U.~Schollw\"ock.
\newblock {The density-matrix renormalization group}.
\newblock \emph{Rev. Mod. Phys.}, 77:\penalty0 259--315, 2005.
\newblock \doi{10.1103/RevModPhys.77.259}.

\bibitem[Schollw\"ock(2011)]{Schollwock2011}
U.~Schollw\"ock.
\newblock {The density-matrix renormalization group in the age of matrix
  product states}.
\newblock \emph{Ann. Phys.}, 326:\penalty0 96--192, 2011.
\newblock \doi{10.1016/j.aop.2010.09.012}.

\bibitem[Shrestha and Mahmood(2019)]{Shrestha2019}
A.~Shrestha and A.~Mahmood.
\newblock {Review of deep learning algorithms and architectures}.
\newblock \emph{IEEE Access}, 7:\penalty0 53040--53065, 2019.
\newblock \doi{10.1109/ACCESS.2019.2912200}.

\bibitem[Sorella(2001)]{Sorella2001}
S.~Sorella.
\newblock {Generalized Lanczos algorithm for variational quantum Monte Carlo}.
\newblock \emph{Phys. Rev. B}, 64:\penalty0 024512, 2001.
\newblock \doi{10.1103/PhysRevB.64.024512}.

\bibitem[Suzuki et~al.(2013)Suzuki, Inoue, and Chakrabarti]{Suzuki2013}
S.~Suzuki, J.~Inoue, and B.~K. Chakrabarti.
\newblock \emph{{Quantum Ising Phases and Transitions in Transverse Ising
  Models}}.
\newblock Springer, Berlin, Heidelberg, 2013.
\newblock ISBN 978-3-642-33039-1.
\newblock \doi{10.1007/978-3-642-33039-1}.

\bibitem[Trabelsi et~al.(2018)Trabelsi, Bilaniuk, Zhang, Serdyuk, Subramanian,
  Santos, Mehri, Rostamzadeh, Bengio, and Pal]{Trabelsi2018}
C.~Trabelsi, O.~Bilaniuk, Y.~Zhang, D.~Serdyuk, S.~Subramanian, J.~F. Santos,
  S.~Mehri, N.~Rostamzadeh, Y.~Bengio, and C.~J. Pal.
\newblock {Deep complex networks}.
\newblock In \emph{International Conference on Learning Representations}, 2018.
\newblock URL \url{https://openreview.net/forum?id=H1T2hmZAb}.

\bibitem[Vapnik(1999)]{Vapnik1999}
V.~N. Vapnik.
\newblock {An overview of statistical learning theory}.
\newblock \emph{IEEE Trans. Neural Netw.}, 10:\penalty0 988--999, 1999.
\newblock \doi{10.1109/72.788640}.

\bibitem[Vidal(2004)]{Vidal2004}
G.~Vidal.
\newblock {Efficient simulation of one-dimensional quantum many-body systems}.
\newblock \emph{Phys. Rev. Lett.}, 93:\penalty0 040502, 2004.
\newblock \doi{10.1103/PhysRevLett.93.040502}.

\bibitem[White and Feiguin(2004)]{WhiteFeiguin2004}
S.~R. White and A.~E. Feiguin.
\newblock {Real-time evolution using the density matrix renormalization group}.
\newblock \emph{Phys. Rev. Lett.}, 93:\penalty0 076401, 2004.
\newblock \doi{10.1103/PhysRevLett.93.076401}.

\end{thebibliography}

\onecolumn\newpage

\appendix

\section{Wirtinger formalism}
\label{sec:Wirtinger}

We employ the Wirtinger formalism to compute gradients of the cost function with respect to complex-valued parameters. This formalism is not widely employed in the context of artificial neural networks, but see e.g.\ \cite{Hirose2012, Trabelsi2018} and references therein. The \emph{Wirtinger} or \emph{Dolbeault} operators are defined as
\begin{equation}
\frac{\partial}{\partial z} := \frac{1}{2} \left(\frac{\partial}{\partial x} - i \frac{\partial}{\partial y}\right), \quad \frac{\partial}{\partial z^*} := \frac{1}{2} \left(\frac{\partial}{\partial x} + i \frac{\partial}{\partial y}\right)
\end{equation}
with $z = x + i y$, $x, y \in \R$. The operators act on (real-) differentiable (identifying $\mathbb{C} \simeq \R^2$) functions $f: U \to \mathbb{C}$ (with $U \subset \mathbb{C}$ some open subset of $\mathbb{C}$), which need not be holomorphic. However, in case $f$ is indeed holomorphic, the Cauchy-Riemann equations imply that the Wirtinger derivative $\partial/\partial z$ is equal to the complex derivative of $f$, whereas the conjugated Wirtinger derivative vanishes:
\begin{equation}
\frac{\partial f(z)}{\partial z} = f'(z), \quad \frac{\partial f(z)}{\partial z^*} = 0 \quad \forall z \in U, \quad f \text{ holomorphic}
\end{equation}
Complex conjugating the second identity leads to $\partial f^*(z)/\partial z = 0$ in this case.

Note that for real-valued functions $f: U \to \R$ (like cost functions considered below), the partial derivatives with respect to $x$ and $y$ can be obtained from the Wirtinger derivative via
\begin{equation}
\label{eq:real_partial_wirtinger}
\frac{\partial f}{\partial x} = 2 \, \mathrm{Re}\!\left(\frac{\partial f}{\partial z}\right), \quad \frac{\partial f}{\partial y} = -2 \, \mathrm{Im}\!\left(\frac{\partial f}{\partial z}\right).
\end{equation}

The following chain rule can be verified by a straightforward calculation:
\begin{equation}
\label{eq:chain_rule_wirtinger}
\frac{\partial}{\partial z} (g \circ f) = \left(\frac{\partial g}{\partial w} \circ f\right) \cdot \frac{\partial f}{\partial z} + \left(\frac{\partial g}{\partial w^*} \circ f\right) \cdot \frac{\partial f^*}{\partial z}.
\end{equation}

The formalism generalizes naturally to functions of several variables; for $z \in \C^n$, we write
\begin{equation}
\nabla^{\text{W}}_z = \left( \frac{\partial}{\partial z_1}, \dots, \frac{\partial}{\partial z_n} \right)^T
\end{equation}
for the Wirtinger nabla operator.

\bigskip

Let $f_{\theta}: U \subset \C^n \to \C^m$ denote the map defined by an artificial neural network with complex-valued parameters $\theta \in \C^p$, input dimension $n$ and output dimension $m$. In abstract terms, our goal is to minimize a cost function with respect to the network parameters via some version of gradient descent (or more precisely, minimizing the expected prediction error for data not used during training \cite{Vapnik1999}):
\begin{equation}
\min_{\theta} C(\theta), \quad C(\theta) = \sum_{j=1}^N c\!\left(f_{\theta}\big(x^{(j)}\big), y^{(j)}\right)
\end{equation}
with $c: \C^m \times \C^k \to \R$ depending on the network output and training labels $y^{(j)} \in \C^k$. Here $(x^{(j)}, y^{(j)})_{j=1,\dots,N}$ a given sequence of training samples. Since $C$ is real-valued, it cannot be holomorphic (except for the trivial case of a constant function), which motivates the use of Wirtinger derivatives in the first place.

Nevertheless, we assume that $f_{\theta}$ is holomorphic as function of the parameters $\theta$. Applying the chain rule \eqref{eq:chain_rule_wirtinger} and using that $\nabla^{\text{W}}_{\theta} f_{\theta}^* = 0$ leads to
\begin{equation}
\label{eq:cost_gradient}
\nabla^{\text{W}}_{\theta} C(\theta) = \sum_{j=1}^N \sum_{k=1}^m \frac{\partial c\!\left(f_{\theta}\big(x^{(j)}\big), y^{(j)}\right)}{\partial f_{\theta,k}\big(x^{(j)}\big)} \, \nabla^{\text{W}}_{\theta} f_{\theta,k}\big(x^{(j)}\big)
\end{equation}
with the subscript $k$ denoting the $k$-th output component of $f_{\theta}$. From Eq.~\eqref{eq:cost_gradient}, one obtains the gradient with respect to the real and imaginary parts of $\theta$ via Eq.~\eqref{eq:real_partial_wirtinger}.

\bigskip

As basic example, the Wirtinger derivative of the quadratic cost (for $a, y \in \C^m$)
\begin{equation}
c(a, y) = \lVert a - y \rVert^2 = \sum_{j=1}^m \lvert a_j - y_j \rvert^2
\end{equation}
reads
\begin{equation}\label{wirtinger_quadratic}
\frac{\partial c(a, y)}{\partial a_j} = (a_j - y_j)^*.
\end{equation}

Combined with the chain rule, the gradients of the cost function in Eq.~\eqref{cost_function} thus read
\begin{equation}
\frac{\partial C}{\partial \theta_{\ell}} %
= \frac{\partial C}{\partial (A\psi)}\frac{\partial (A\psi)}{\partial \psi}\frac{\partial \psi}{\partial \theta_{\ell}} %
= \left\langle A\psi - b \Big\vert A \frac{\partial \psi}{\partial \theta_{\ell}} \right\rangle
\end{equation}

\section{Error analysis}
We distinguish between three different wave functions: the exact one, $\psi(t)$; the one obtained by the exact midpoint time-evolution, $\psi_{\Delta}(t)$; and the one represented by the network, $\psi_N(t)$. We want to find the error of the network with respect to the exact state:
\begin{equation}
    \varepsilon(t) = \psi(t) - \psi_N(t)
\end{equation}
This can be split into the error due to the the midpoint method and the error due to the network optimization:
\begin{equation}
    \varepsilon(t) = \varepsilon_{\Delta}(t) + \varepsilon_N(t)
\end{equation}
with
\begin{equation}
    \varepsilon_{\Delta}(t) = \psi(t) - \psi_{\Delta}(t)
\end{equation}
and
\begin{equation}
    \varepsilon_N(t) = \psi_{\Delta}(t) - \psi_N(t).
\end{equation}
Using the triangle inequality we can set an upper bound to the absolute error:
\begin{equation}
    | \varepsilon(t) | \leq | \varepsilon_{\Delta}(t) | + | \varepsilon_N(t) |
\end{equation}
As implied by Eq.~\eqref{midpoint}, in order to obtain the state at the next time step using the midpoint rule one must solve the following linear matrix equation:
\begin{equation}\label{exactmid}
    A \psi_{\Delta}( t_{n + 1} ) - B \psi_{\Delta}(t_n) = 0
\end{equation}
where
\begin{equation}
    A = I + \frac{i \Delta t}{2}H
\end{equation}
and
\begin{equation}
    B = I - \frac{i \Delta t}{2}H.
\end{equation}
As a second-order Runge-Kutta method, its global error is of the order of $O(\Delta t^2)$. However, instead of solving this exactly, our method optimizes $\psi_{N} (t_{n+1})$ to minimize the above. That means we are actually solving the equation
\begin{equation}\label{oursys}
    A \psi_{N}( t_{n + 1} ) - B \psi_{N}(t_n) = r^{(n+1)},
\end{equation}
where $r^{(n+1)}$ is the residual obtained at the end of the optimization in time-step $n+1$. We are interested in finding an expression for $\varepsilon_N(t_n)$. Using its definition into Eq.~\eqref{oursys} we obtain
\begin{equation}
    A(\psi_{\Delta}(t_{n+1}) - \varepsilon_N(t_{n+1})) - B(\psi_{\Delta}(t_{n}) - \varepsilon_N(t_{n})) = r^{(n+1)},
\end{equation}
and since the exact midpoint rule must be 0 (see Eq.~\eqref{exactmid}) this simplifies to
\begin{equation}
    \varepsilon_N(t_{n+1}) = A^{-1}\left(B\varepsilon_N(t_n) - r^{(n+1)}\right).
\end{equation}
Similarly as what we did for the midpoint error, we now have a recursive relation for the network error for the next time step. Then, by proof of induction one can show that
\begin{equation}
    \varepsilon_N (t_n) = -B^{-1}\sum_{m = 1}^{n} A^{-m}B^{m} r^{(n + 1 - m)},
\end{equation}
for all $n > 0$. Note that the application of $A^{-1}B$ corresponds to one time step of the midpoint time evolution. This means that the residual from each time step is added to the error and time-evolved unitarily, so the residual will not dramatically increase the error as an artifact of the chosen integration method.

\section{Effect of time step size, and comparison with the forward Euler method}

Fig.~\ref{fig:comparison} in the main text directly compares SR with the midpoint optimization using the same number of time steps. However, the implicit midpoint integration method is a second-order method, which suggests the possibility of achieving a similar final error using a larger time step size for the real time evolution. Here we consider the effects of varying the time step size. Furthermore, we compare the results with an analogous neural network optimization based on the forward Euler method.

\begin{figure}
    \centering
    \includegraphics[width=0.5\columnwidth]{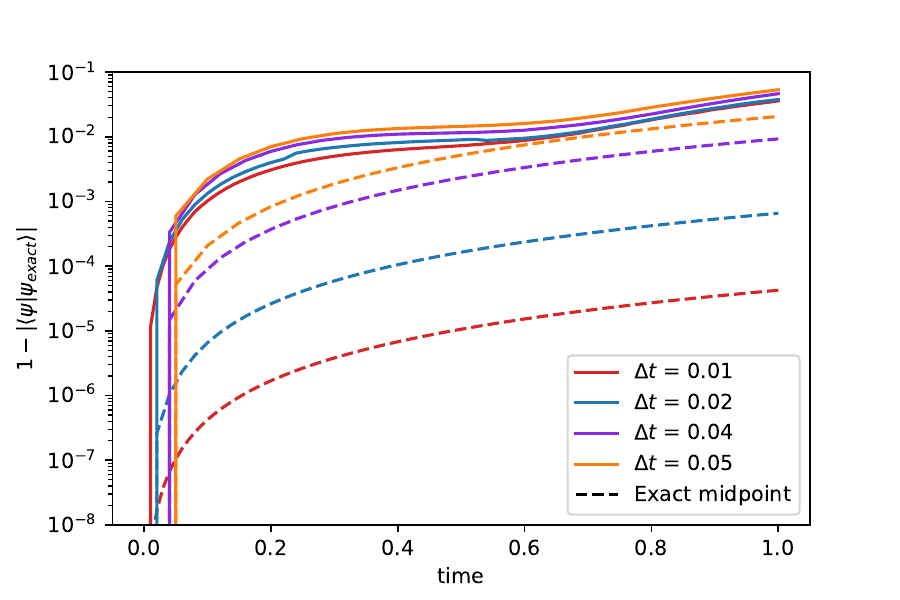}
    \caption{Overlap error achieved with different time steps for the same system under the same optimization hyperparameters (RBM with 60 hidden units, same learning rate and same number of optimization steps).}
    \label{fig:timestep}
\end{figure}

We apply the midpoint optimization to simulate a quench from $h = 1.5$ to $h = 0.75$ for a one-dimensional system with 15 lattice sites. As shown in Fig.~\ref{fig:timestep} the overlap error resulting from the optimization shows very little variation as one decreases $\Delta t$, different from an exact midpoint integration. These results further strengthen our intuition that most of the error originates from the expressiveness of the architecture, and not the integration method. The variation that is seen can be explained by the fact that the same learning rate and number of optimization parameters are being used in each simulation. For a larger $\Delta t$ one can expect a larger update in the network parameters, and a longer optimization may be necessary. In practice, the small improvement may not be worth the computational expense of decreasing $\Delta t$ nor increasing the number of optimization steps.

\begin{figure*}[ht!]
\centering
\subfloat[Overlap error]{
    \includegraphics[width=0.5\columnwidth]{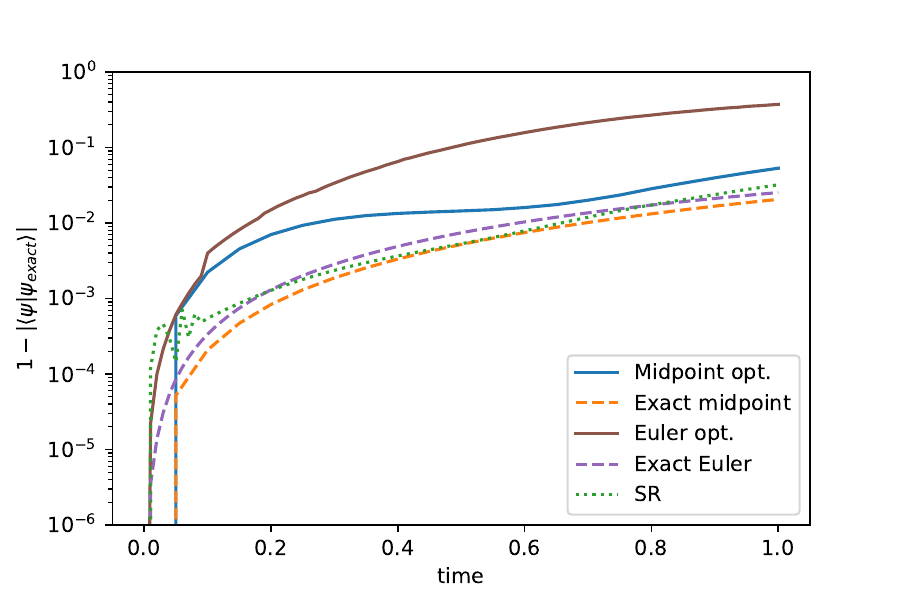}
    \label{fig:euler_comparison}}
\subfloat[Example of loss convergence at $t=0.5$]{
    \includegraphics[width=0.5\columnwidth]{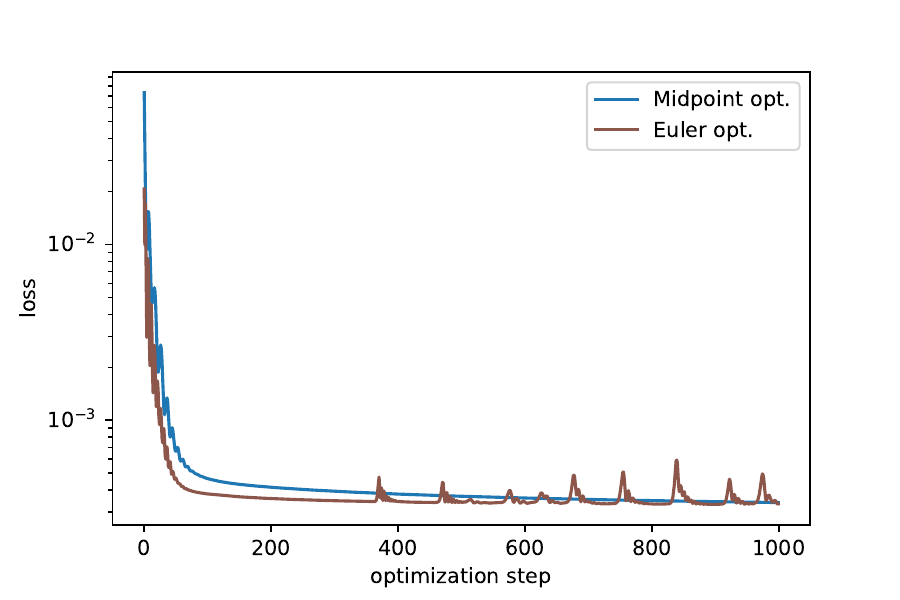}
    \label{fig:losses}}
\caption{Comparison of different integration methods and time steps: SR and the Euler method with $\Delta t = 0.01$ and the midpoint method with $\Delta t = 0.05$. In (b) the loss is normalized with respect to the number of samples and time steps.}
\end{figure*}

Next we compare these results with those obtained using an Euler step for the optimization. In a similar fashion to Eq.~\eqref{cost_sampling} we choose as cost function
\begin{equation}
C(\theta_{n+1}) = \sum_{j=1}^N \left\vert \left( \psi[\theta_{n+1}] - i \Delta t H \ \psi[\theta_n] \right)\left(\sigma^{(j)} \right) \right\vert^2.
\end{equation}
Fig.~\ref{fig:euler_comparison} shows that the results obtained are worse than those obtained with the implicit midpoint rule, despite the similar errors of the exact Euler and midpoint integrations for the chosen $\Delta t$ sizes. Taking a closer look at the optimization landscape shows that the loss exhibits a smoother convergence with the midpoint optimization, as shown in Fig.~\ref{fig:losses}. Implementing an early stop mechanism may be beneficial for the Euler optimization.

\section{Geometrical perspective}
\label{sec:Geometrical}

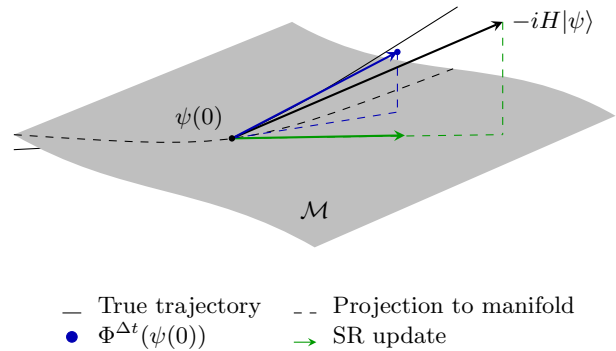
\begin{figure}[!ht]
\centering
\begin{tikzpicture}[>=stealth]

\path[name path=border1] (0,0) to[out=-30,in=150] (3.7,1.5);
\path[name path=border2] (8,0.25) to[out=150,in=-30] (3.7,1.5);
\path[name path=redline] (0,0) -- (8,1.2);
\path[name intersections={of=border1 and redline,by={a}}];
\path[name intersections={of=border2 and redline,by={b}}];

\coordinate (O) at (2.9,-0.05);

\draw[black] (0, -0.2) -- (O);

\fill[color=gray!50] 
  (0,0) to[out=-30,in=150] (4,-1.5) -- (8,0.25) to[out=150,in=-30] (3.7,1.5) -- cycle;
\node at (4,-1) {$\mathcal{M}$};

\draw[black] (O) .. controls (4, 0.5) .. (5.9,1.7);

\draw[black, dashed] 
  (a) .. controls (3,-0.2) .. (b);

\draw[->, thick] (O) -- (6.5,1.5) node[anchor=west] {$-iH \lvert \psi \rangle$};

\draw[color=black!40!green, dashed] (6.5,1.5) -- (6.5,0);
\draw[color=black!40!green, dashed] (O) -- (6.5,0);

\draw[->, color=black!40!green, thick] (O) -- (5.2, -0.01);

\draw[->, thick, color=black!30!blue] (O) -- (5.1, 1.1);
\draw[color=black!30!blue, dashed] (5.1, 1.1) -- (5.1, 0.3);
\draw[color=black!30!blue, dashed] (O) -- (5.1, 0.3);
\draw[color=black!30!blue, fill] (5.1, 1.1) circle (1pt);

\filldraw[black] (O) circle (1pt) node[anchor=south east] {$\psi(0)$};

\node at ([yshift=-1cm]current bounding box.south) {
\setlength\tabcolsep{3pt}
\begin{tabular}{@{}cl@{\hspace{12pt}}cl@{}}
\tikz\draw[black](0,0)--(.25,0); & True trajectory & \tikz\draw[color=black, dashed](0,0)--(.3,0); & Projection to manifold \\ 
\tikz\draw[color=black!30!blue, fill] (0,0) circle (2pt); & $\Phi^{\Delta t}(\psi(0))$ 
  & \tikz\draw[color=black!40!green, ->] (0,0) -- (0.3,0); & SR update
\end{tabular}
};
\end{tikzpicture}
\caption{Geometric interpretation of the update achieved by SR and a gradient descent based method from a numerical integration flow $\Phi^{\Delta t}$}
\label{fig:geometricfig}
\end{figure}

From a geometrical perspective, the architecture of the neural network defines the manifold $\mathcal{M}$ to which the time-evolution is restrained. As shown in Fig.~\ref{fig:geometricfig}, in SR the parameters are updated by projecting the true derivative at the current time onto $\mathcal{M}$. Similarly, the minimization of Eq.~\eqref{minimization}, implicitly projects the next timestep obtained from the integration flow $\Phi$, as the point of minimum distance in the manifold is its projection.

\section{Details of stochastic reconfiguration calculation}

\begin{figure}[h!]
\centering
\subfloat[Effect of cut-off]{\includegraphics[width=0.5\linewidth]{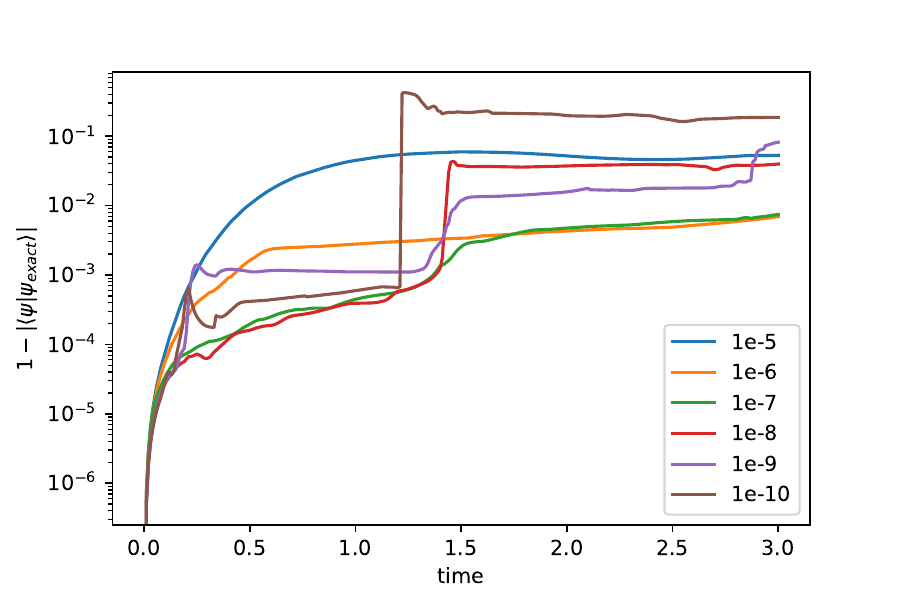}
\label{pseudo}}
\subfloat[Comparison of solvers]{\includegraphics[width=0.5\linewidth]{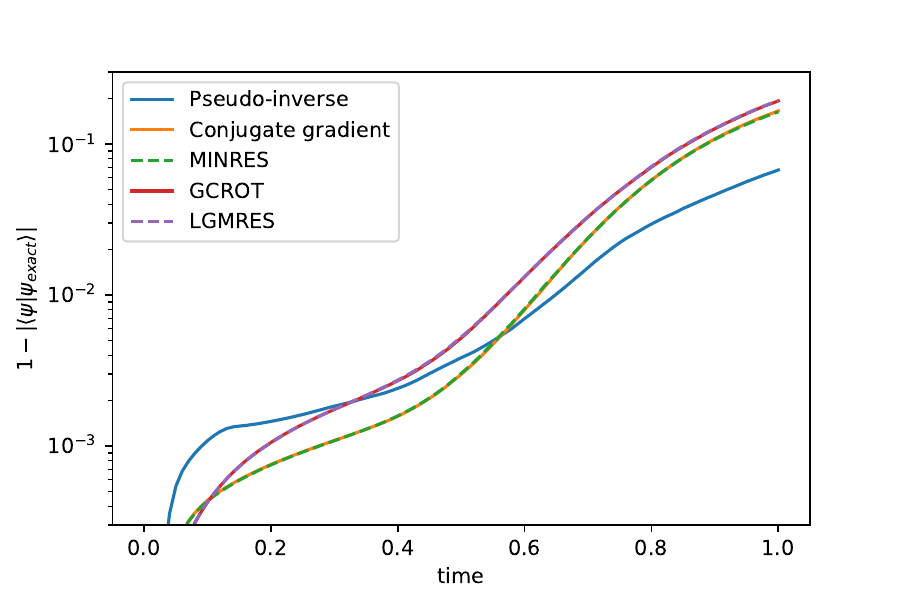}
\label{fig:iterative}}
\caption{Comparison in error using SR with (a) different cut-offs for the pseudo-inverse and (b) different iterative solvers, as well as using a pseudo-inverse with cut-off at $10^{-10}$.}
\end{figure}

Fig.~\ref{fig:comparison} in the main text shows a comparison between our method and SR. For our simulations, we determine the cut-off of the pseudo-inverse as follows: we perform the time evolution several times decreasing the cut-off by one order of magnitude at each run, from $0.1$ to $10^{-11}$, and chose the one that resulted in the smallest error at $t = 1$. It is worth noting, however, that the best cut-off at some time $t_1$ may not be the best at a later time $t_2$. Fig.~\ref{pseudo} illustrates this point for an Ising chain with 6 lattice sites.

We also tried several Krylov subspace methods for solving Eq.~\eqref{eq:Stoch_rec}, but found their performance to be worse than using the pseudo-inverse with optimal threshold, as shown in Fig.~\ref{fig:iterative}.

\section{Filter weights visualization}

\begin{figure}
\centering
\subfloat[Amplitudes of filter weights]{\includegraphics[width=0.5\columnwidth]{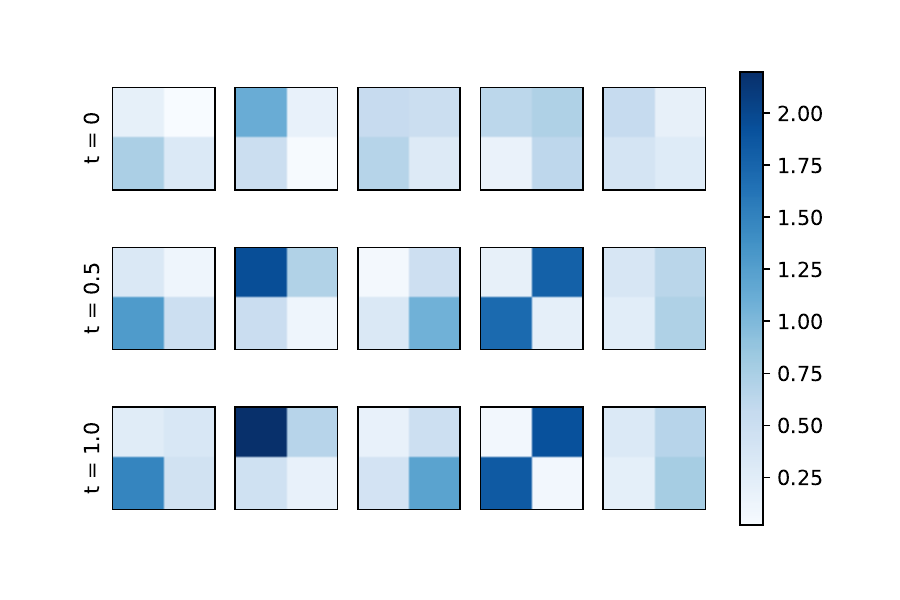}}
\subfloat[Arguments of filter weights]{\includegraphics[width=0.5\columnwidth]{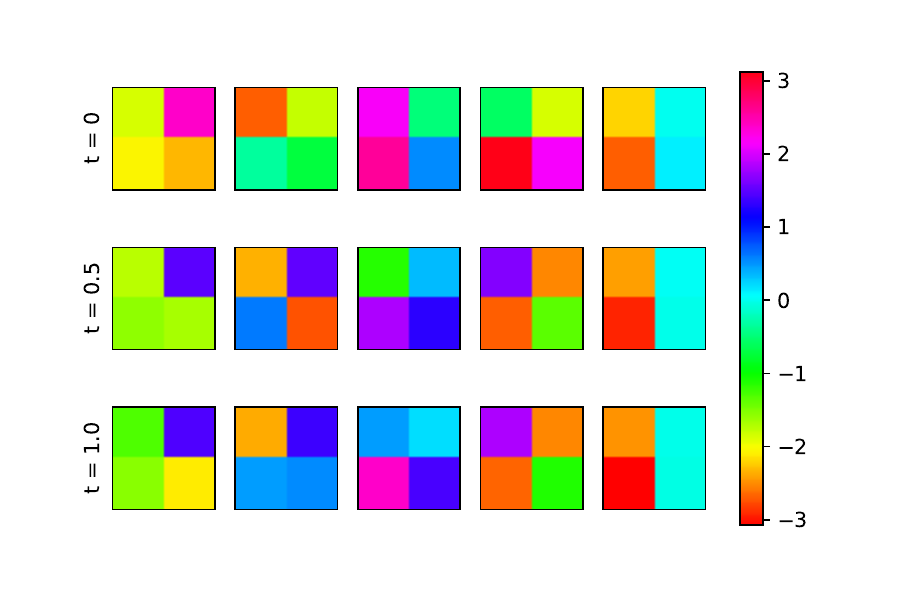}}
\caption{(a) Amplitudes and (b) arguments of the complex weights for each filter of the CNN in Fig.~\ref{fig:CNN}, for the $h = 2 h_c$ simulation shown in Fig.~\ref{fig:2d_tevol}. The rows corresponds to different time points.}
\label{fig:filters}
\end{figure}

As illustration, Fig.~\ref{fig:filters} shows the five complex-valued convolution layer filter weights for the time evolution of the 2D system at three time points, for the quench to $h = 2 h_c$. It appears that the amplitudes vary rather slowly as compared to the complex arguments (phases).

\end{document}